\documentclass[3p]{elsarticle}

\usepackage{hyperref}
\usepackage{amssymb}
\usepackage{bm}
\usepackage{dcolumn}
\usepackage{amssymb}
\usepackage{amsmath}
\usepackage{graphicx}

\usepackage{epstopdf}
\usepackage{epsfig}

\usepackage{subfig}
\usepackage{booktabs}
\usepackage{multirow}
\usepackage{setspace}
\usepackage{array}
\usepackage{threeparttable}
\usepackage{txfonts}

\usepackage{exscale}
\usepackage{relsize}
\journal{Phys. Lett. B 820 (2021) 136518, \url{https://doi.org/10.1016/j.physletb.2021.136518}}

\begin{document}

\begin{frontmatter}



  \newcommand*{\PKU}{School of Physics, Peking University, Beijing 100871,
    China}
  \newcommand*{\CHEP}{Center for High Energy Physics, Peking University, Beijing 100871, China}
   \newcommand*{\CIC}{Collaborative Innovation Center of Quantum Matter, Beijing, China}

  \title{Pre-burst events of gamma-ray bursts with light speed variation}

  \author[a]{Jie Zhu}
  \author[a,b,c]{Bo-Qiang Ma\corref{cor1}}

  \address[a]{\PKU}
  \address[b]{\CHEP}
  \address[c]{\CIC}

%  \address[d]{\CHPS}
  \cortext[cor1]{Corresponding author \ead{mabq@pku.edu.cn}}

  \begin{abstract}

    Previous researches on high-energy photon events from gamma-ray bursts~(GRBs) suggest a light speed  variation
    $v(E)=c(1-E/E_{\mathrm{LV}})$ with $E_{\mathrm{LV}}=3.6\times10^{17}~\mathrm{ GeV}$, together with a pre-burst scenario that hight-energy photons
    come out about 10 seconds earlier than low-energy photons at the GRB source. However, in the Lorentz invariance violating scenario with an energy dependent light speed
    considered here, high-energy photons travel slower than low-energy photons due to the
    light speed variation, so that they are usually detected after low-energy photons in observed GRB data.
    Here we find four high-energy photon events which
    were observed earlier than low-energy photons from Fermi Gamma-ray Space Telescope~(FGST), and analysis on these photon events supports the pre-burst scenario
    of high energy photons from GRBs and the energy dependence of light speed listed above.

  \end{abstract}


\end{frontmatter}

\section{Introduction}

According to Einstein's relativity, light speed is a constant $c$ in free space. However, it is speculated from quantum gravity
that the Lorentz invariance might be broken at the Planck scale~($E_\mathrm{Pl}\simeq 1.22\times10^{19}~\mathrm{GeV}$),
and that the light speed may have a variation with the energy of the photon. Amelino-Camelia {\it et al.}~\cite{method1, method2} first suggested testing
Lorentz violation by comparing the arrival times between high energy and low energy photons from gamma-ray bursts~(GRBs).
For energy $E\ll E_{\rm Pl}$, the modified dispersion
relation of the photon can be expressed in leading order as
\begin{equation}\label{eq:1}
  E^2=p^2 c^2 \left[1-s_n\left(\frac{pc}{E_{\mathrm{LV,} n}}\right)^n\right].
\end{equation}
Assuming that the traditional relation $v=\partial E / \partial p$ holds, we have the following speed relation
\begin{equation}\label{eq:2}
  v(E)=c\left[1-s_n\frac{n+1}{2}\left(\frac{pc}{E_{\mathrm{LV,}n}}\right)^n\right],
\end{equation}
where $n=1$ or $n=2$ as usually assumed, $s_n=\pm1$ indicates whether high-energy photons travel faster~($s_n=-1$)
or slower~($s_n=+1$) than low-energy photons, and $E_{\rm{LV},n}$ represents the nth-order Lorentz violation scale.
From Eq.~(\ref{eq:2}) we can derive a time lag between two photons with different energies in flat universe, however
we need to consider the expansion of the Universe~\cite{formula}, and the result shows as
\begin{equation}\label{eq:3}
  \Delta t_{\mathrm{LV}}=s_n\frac{1+n}{2H_0}\frac{E^n_{\mathrm{h}}-E^n_{\mathrm{l}}}{E^n_{\mathrm{LV,}n}}\int_0^z\frac{(1+z')^n\mathrm{d}z'}
  {\sqrt{\Omega_{\mathrm{m}}(1+z')^3+\Omega_{\Lambda}}},
\end{equation}
where $E_{\rm{h}}$ and $E_{\rm{l}}$ correspond to the energies of the observed high-energy and
low-energy photons, $z$ is the redshift of the source GRB, $H_0$, $\Omega_{\rm{m}}$ and $\Omega_{\rm{\Lambda}}$
are cosmological constants. Here we adopt the present day Hubble constant $H_0=67.3\pm 1.2~ \rm{km ~s}^{-1}\rm{Mpc}^{-1}$~\cite{pgb},
the pressureless matter density $\Omega_{\rm{m}}=\mathrm{0.315^{+0.016}_{-0.017}}$~\cite{pgb} and the dark energy density
$\Omega_{\Lambda}=\mathrm{0.685^{+0.017}_{-0.016}}$~\cite{pgb}.

The observed time difference between high energy and low energy photons should not be only the time lag due to Lorentz violation,
i.e., Eq.(\ref{eq:3}), but also an intrinsic time lag $\Delta t_{\rm{in}}$ at the source GRB~\cite{Ellis,shaolijing,zhangshu,xu1,xu2}, which means that in the source
reference system high-energy photon events and low-energy photon events have an intrinsic time difference $\Delta t_{\rm{in}}$.
Considering the expansion of the Universe, we have
\begin{equation}\label{eq:4}
  \Delta t_{\mathrm{obs}}=\Delta t_{\mathrm{LV}}+(1+z) \Delta t_{\mathrm{in}},
\end{equation}
where $\Delta t_{\mathrm{obs}}$ is the difference of observed arrival times between high-energy and
low-energy photons, $z$ is the redshift of the source GRB and $\Delta t_{\mathrm{LV}}$ is the time lag caused by
Lorentz violation as expressed in Eq.~(\ref{eq:3}).
In fact, with cosmic photons from one single source, one has difficult to make clear distinction between the Lorentz violation effect and the intrinsic source effect from the observed time difference $\Delta t_{\mathrm{obs}}$, see Eq.~(\ref{eq:4}). The combination of multi-GeV photons from GRBs with different redshifts renders it feasible to make distinction
between Lorentz violation effect and intrinsic source effect.

Previous studies~\cite{xu1,xu2} on high energy photon events from GRBs detected by Fermi Gamma-ray Space Telescope (FGST)~\cite{LAT,GBM}
suggest a regularity of high energy photon events
with a conclusion that $s=+1$, $n=1$, $E_{\rm{LV, }1}=(3.60\pm0.26)\times10^{17}~\rm{GeV}$ and
$\Delta t_{\rm{in}}=(-10.7\pm1.5)~\rm{s}$. In this physics picture, it is suggested that high-energy photons
come out about 10 seconds earlier than low-energy photons at the GRB source, and because of light speed variation, high-energy photons
travel slower than low-energy photons, and the light speed difference and the long cosmological distances lead to an expectation
that one usually observes low-energy photons earlier than high-energy photons, as is indeed the case in the earlier observations
of GRB data.

\begin{figure}
  \centering
  %\includegraphics[width=90mm]{condition.eps}\\
  \includegraphics[width=90mm]{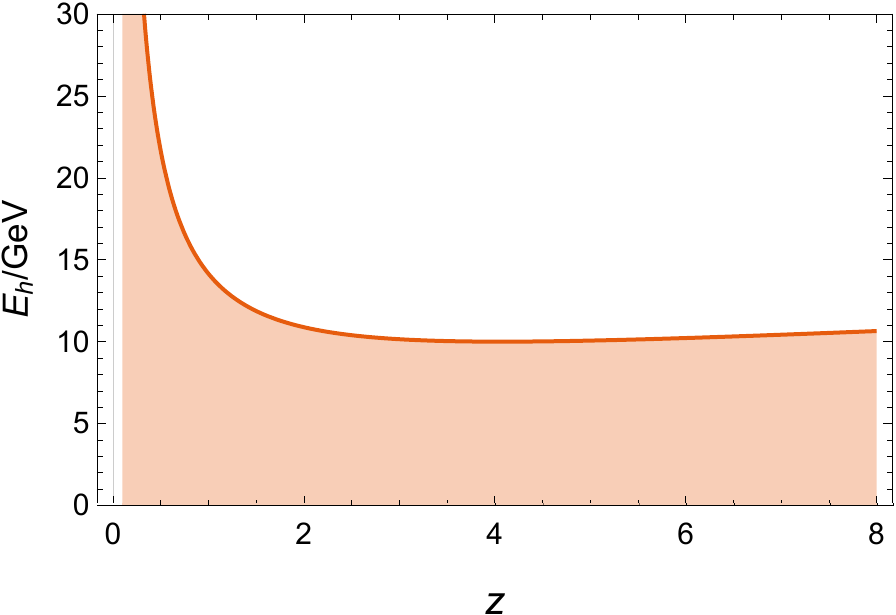}\\
  \caption{Plot of the condition Eq.~(\ref{eq:6}) to observe pre-burst events. $z$ is the redshift of the GRB source, and $E_{\rm{h}}$ is
    the observed energy of high-energy photon event. The shaded part represents the area of Eq.~(\ref{eq:6}) }
  \label{fig:condition}
\end{figure}

Here we want to search for high-energy photon events which are observed earlier than low-energy ones,
and we call these events as \textbf{observed pre-burst events}. In traditional
point of view one may consider these events as just background noises without any significance, but in the picture of light speed variation, these events should
come from pre-burst emission of high energy photons from GRBs so that they are novel signals to support the observed regularity
as an indication for the light speed variation~\cite{xu1,xu2}. However to find observed pre-burst events is not easy. If the energy of the photon is too high, the speed
of the photon may cause it fell behind low-energy photons. Just take the conclusion of Refs.~\cite{xu1,xu2} and do a simple
calculation, if we want to find observed pre-burst events, we have
\begin{equation}\label{eq:5}
  \Delta t_{\mathrm{LV}}+(1+z) \Delta t_{\mathrm{in}}<0.
\end{equation}
Usually $E_{\rm{h}}\sim \rm{GeV}$ while $E_{\rm{l}}\sim \rm{keV}$, so it is reasonable
to take $E_{\rm{l}}$ as 0.
Combining Eq.~(\ref{eq:5}) with Eq.~(\ref{eq:3}) and let $n=1$, we have
\begin{equation}\label{eq:6}
  E_{\rm{h}}<- E_{\rm{LV, }1} H_0 \Delta t_{\mathrm{in}}  (1+z)/ \int_0^z\frac{(1+z')\mathrm{d}z'}
  {\sqrt{\Omega_{\mathrm{m}}(1+z')^3+\Omega_{\Lambda}}}.
\end{equation}
As shown in Fig.~\ref{fig:condition}, we can not expect that we could find observed pre-burst events with too high energy.
For example, if the redshift of a GRB is 2, we can only expect observed pre-burst high-energy photons with energy
less than $10.9~\rm{GeV}$.

\section{Data Acquirement}

We search for the data from the Fermi Gamma-ray Space Telescope~(FGST). FGST consists of the Fermi Large
Area Telescope~(LAT)~\cite{LAT} and the Gamma-Ray Burst Monitor~(GBM)~\cite{GBM}. LAT aims to collect high-energy events while
GBM aims to collect low-energy events. The GBM data can be downloaded from the Fermi website~\cite{gbm_data} while the LAT data need to be
retrieved and downloaded from this website~\cite{lat_data}. The trigger time $t_{\rm{trig}}$ of GBM is usually assumed as the onset
time of GRBs, so we search for high-energy events before $t_{\rm{trig}}$. The retrieval scope of time is
set from $t_{\rm{trig}}-50~\rm{s}$ to $t_{\rm{trig}}-0~\rm{s}$, and the retrieval radius is set to 2 degrees.
The energy range is set from 100 MeV to 100 GeV, and the lower limit
is chosen to reject events with poorly reconstructed directions and energies.
%Due to the expansion of the Universe, energy $E_{\rm{obs}}$ measured today corresponds to the energy at the source as
%\begin{equation}\label{eq:7}
%  E_{\rm{source}}=(1+z)E_{\rm{obs}}.
%\end{equation}
%We search for events whose source energy is higher than 1~GeV, so the retrieval scope is set from
%$1/(1+z)~\rm{GeV}$ to maximum.
We have searched 48 GRBs, from GRB080916C to GRB210204A, which are not only detected by FGST but also have redshifts recorded. Although our retrieval scope of time and energy is too much larger than we need, we find only
3 GRBs which have observed pre-burst events recorded, and they are \textbf{GRB~201020A}, \textbf{GRB~201020B} and \textbf{GRB~201021C}.

\begin{figure}[!h]
  \centering
  \subfloat[]{
    \label{fig:1a}
    \begin{minipage}{0.5\textwidth}
      \centering
      \includegraphics[width=\textwidth]{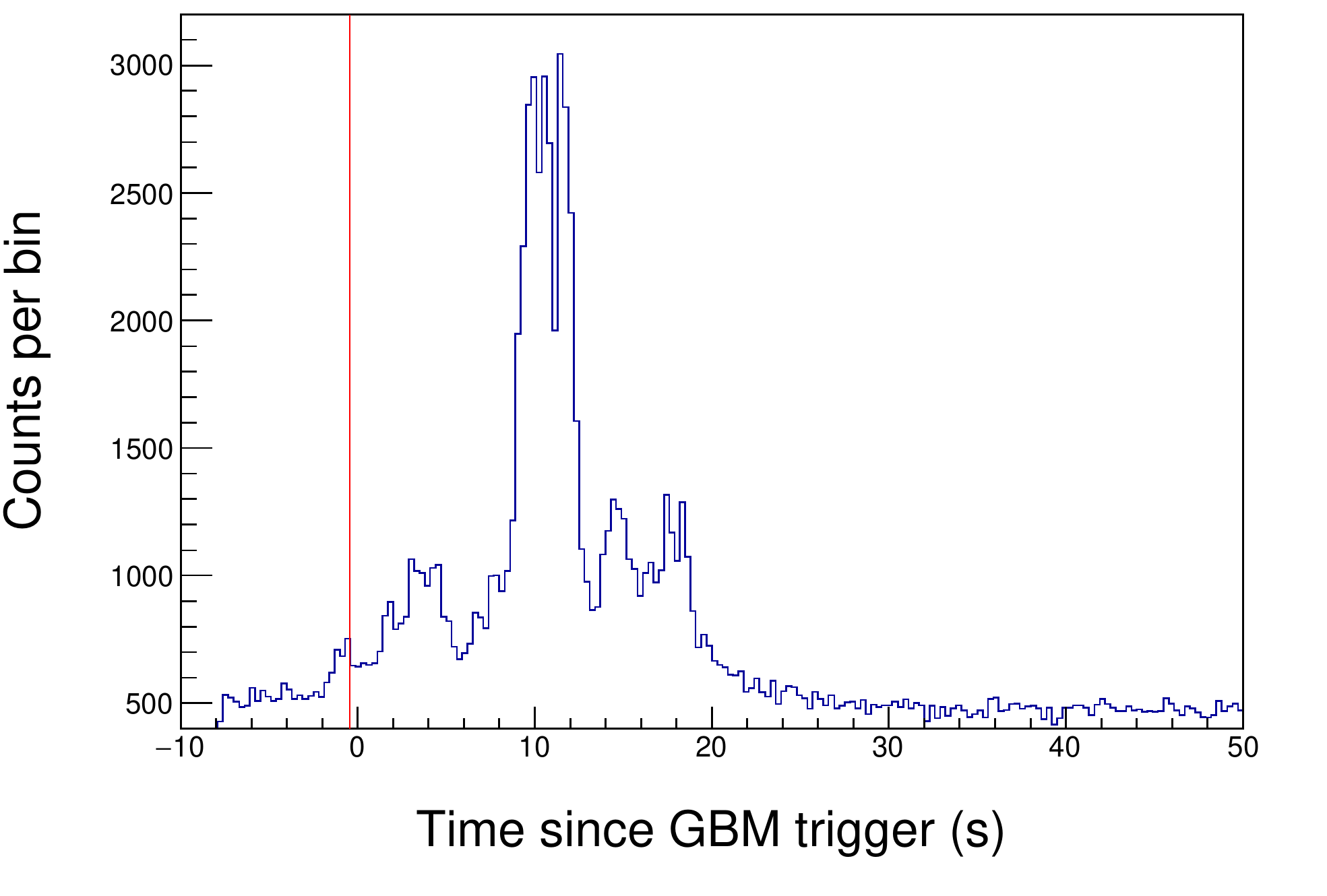}
    \end{minipage}
  }
  \subfloat[]{
    \label{fig:1b}
    \begin{minipage}{0.5\textwidth}
      \centering
      \includegraphics[width=\textwidth]{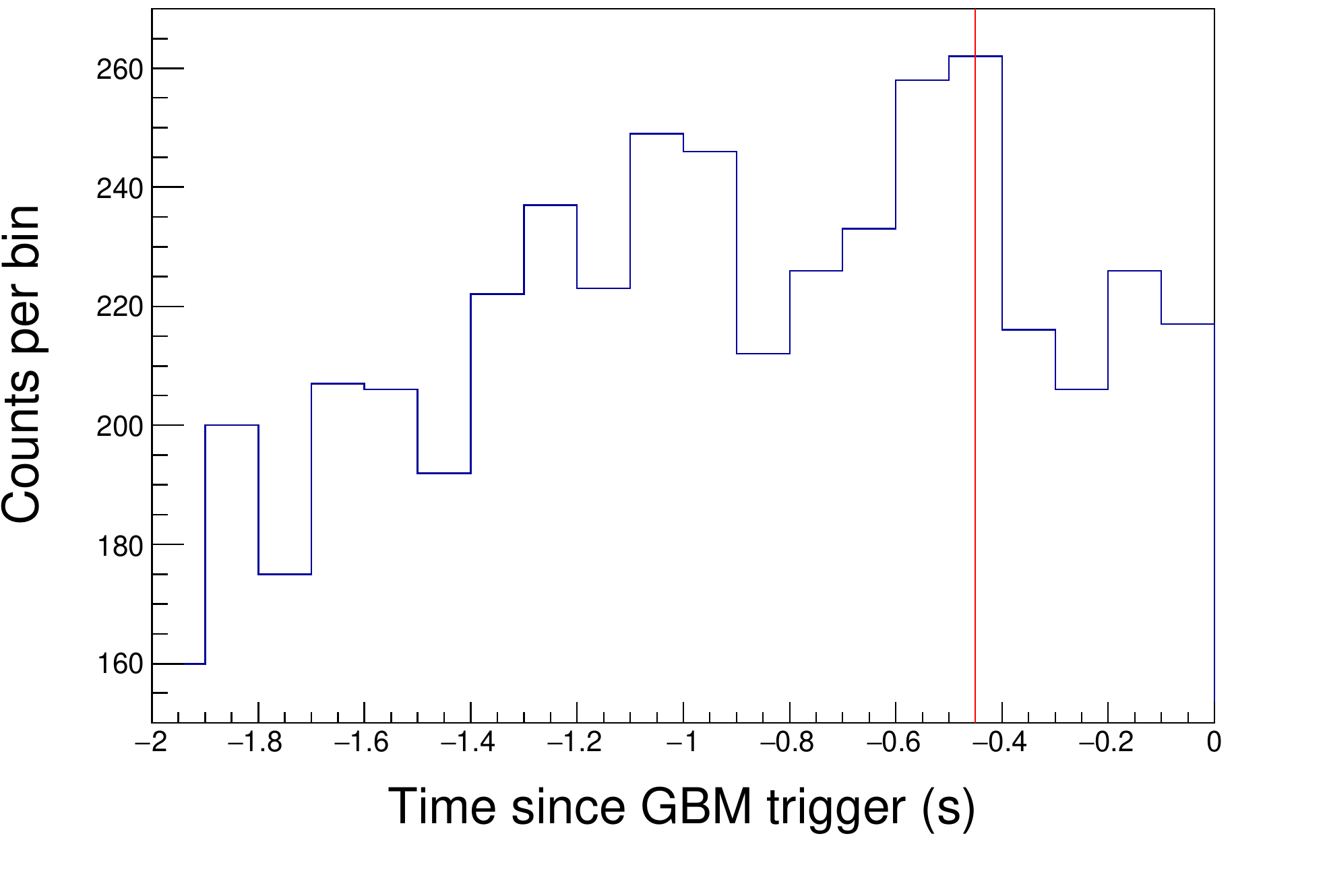}
    \end{minipage}
  }
  \caption{Light curves of the two brightest trigger detectors combined~(GBM NaI-n7 and NaI-nb,
    $7.9\sim 258~\mathrm{keV}$) for GRB~201020B. In the left panel~(a), photon events are binned in
    0.3~second intervals. In the right panel~(b), photon events are binned in 0.1~second intervals
    to determine the peak of the first significant peak as $t_{\rm{low}}=-0.45~\rm{s}$ \label{fig:low}.}
\end{figure}

Here we discuss $\Delta t_{\rm{obs}}$, the difference of observed arrival time between high-energy photon and
low energy photon. In the following discussion, we set $t_{\rm{trig}}$ as the origin of the time coordinate. As defined
\begin{equation}\label{eq:8}
  \Delta t_{\rm{obs}}=t_{\rm{high}}-t_{\rm{low}},
\end{equation}
where $t_{\rm{high}}$ is the
arrival time of high-energy photon and $t_{\rm{low}}$ represents the arrival time of low-energy photon. For
$t_{\rm{high}}$ of single photons, we can get it from the retrieved LAT data, and for $t_{\rm{low}}$, we adopt
the first significant peak criteria discussed by the work of Ref.~\cite{Liu}.
For every GRB in this work, the first significant peak time is close to the trigger time.
For example the first significant peak for GRB201020B indicates that $t_{\rm{low}}=-0.45~\rm{s}$  as shown in
Fig.~\ref{fig:low}.
We list all of the observed pre-burst events of the two GRBs in Table~\ref{tab:events}.

\begin{threeparttable}[t]
  \begin{centering}
    \caption{The data of observed pre-burst events}
    \begin{tabular}{cccccp{24mm}<{\centering}cp{24mm}<{\centering}cp{10mm}<{\centering}}
      \hline
      \hline
      GRB        & $z$   & $t_{\rm high}$~(s) & $t_{\rm low}$~(s) & $E_{\rm obs}$~(GeV) & $E_{\rm source}$~(GeV) & RA($^{\circ}$) & Dec($^{\circ}$) & $p$ \\
      \hline
      201020A    & 2.903 & -22.955            &-1.875             & 0.229               & 0.892                  &-22.955  & 260.264  & 0.79\\
      \hline
      201020B    & 0.804 & -17.405            & -0.45             & 1.227               & 2.214                  & 67.360         & 77.323   & 0.64       \\
      \hline
      201021C(1) & 1.07  & -12.839             & -0.625            & 0.469               & 0.972                  & 11.485        & -54.462  & 0.84        \\
      201021C(2) & 1.07  & -37.173             & -0.625            & 0.155               & 0.322                  & 15.070        & -54.820  & 0.76        \\
      \hline
      \hline
    \end{tabular}%

    \begin{tablenotes}
      \item Data of the observed pre-burst events of the three GRBs. $z$ is the redshift of the GRB,
      $t_{\rm{high}}$ is the arrival time of high-energy photon and $t_{\rm{low}}$ represents
      the first significant peak time of low-energy photons. $E_{\rm{obs}}$ is the energy observed
      by LAT while $E_{\rm{source}}=(1+z)E_{\rm{obs}}$ is the corresponding energy at the source. $\rm{(RA, Dec)}$ is
      the position of the events~(J2000). $p$ is the probability that the event is associated with the source and is generated by the \emph{Fermi} ScienceTool \emph{gtsrcprob}~\cite{ScienceTool}.
      Redshift of GRB~201020A is from Ref.~\cite{201020Az}, redshift of GRB~201020B is from Ref.~\cite{201020Bz} and redshift of GRB~201021C is from Ref.~\cite{201021Cz}.
    \end{tablenotes}
    \label{tab:events}%
  \end{centering}
\end{threeparttable}%
\\

\section{Data Analysis and Result}

Here we use the similar analyse method introduced in Ref.~\cite{xu1}. As a brief introduction, combining Eq.~(\ref{eq:3})
and Eq.~(\ref{eq:4}), we have
\begin{equation}\label{eq:reexpress}
  \frac{\Delta t_{\mathrm{obs}}}{1+z}=s_n \frac{K_n}{E^n_{\mathrm{LV,}n}}+\Delta t_{\mathrm{in}},
\end{equation}
where $K_n$ is the Lorentz violation factor
\begin{equation}\label{eq:LVfactor}
  K_{n}=\frac{1+n}{2H_0}\frac{E^n_{\mathrm{h}}}{1+z}\int_0^z\frac{(1+z')^n\mathrm{d}z'}
  {\sqrt{\Omega_{\mathrm{m}}(1+z')^3+\Omega_{\Lambda}}}.
\end{equation}
Then we make a $\Delta t_{\rm{obs}}/(1+z)$ versus $K_1$ plot and try to find linear relation between different
events.
These photons with a same intrinsic time lag would fall on an inclined straight line in the $\Delta t_{\mathrm{obs}}/(1+z)$~-~$K_{n}$ plot, and we can determine $\Delta t_{\mathrm{in}}$ of them as the intercept of the line with the $Y$ axis. The slope of the mainline is $1/E_{\rm
LV,1}$, from which one can determine the Lorentz violation scale  $E_{\rm LV,1}$.
%%Events falling on another line in parallel with the mainline can be considered as with a same intrinsic time lag different from that of the mainline.
However, it is not reasonable to assume that all of the high-energy photons emit at exactly a same time and that the intrinsic time lag $\Delta t_{\rm{in}}$ for every GRB is the same, and there may be a distribution for the intrinsic time lag $\Delta t_{\rm{in}}$.
Since we know little about the intrinsic emission mechanism of GRBs, here we just assume that $\Delta t_{\rm{in}}$ follows a normal distribution with mean $\mu$ and standard deviation $\sigma$. Here we use Bayesian analysis and maximum likelihood estimation~(MLE) to fit the data~\cite{Ellis,HESS}.
For a linear model $y=kx+b$ and given $n$ groups of data $(x_i, y_i)$ with errors $\sigma_{x_i}$ and $\sigma_{y_i}$, the likelihood function for an individual point can be expressed as
\begin{equation}
  p(y_i|x_i,\sigma_{x_i},\sigma_{y_i},k,b)=\frac{1}{\sqrt{2\pi(\sigma_{x_i}^2k^2+\sigma_{y_i}^2)}}\exp{\left(-\frac{(k x_i+b-y_i)^2}{2(\sigma_{x_i}^2k^2+\sigma_{y_i}^2)}\right)}.
\end{equation}
If the parameter $b$ follows a Gaussian distribution with mean $\mu$ and standard deviation $\sigma$, we can derive the likelihood function for an individual point as
\begin{equation}\label{eq:deviation}
\begin{split}
p(y_i|x_i,\sigma_{x_i},\sigma_{y_i},k,\mu,\sigma) &=\int_{-\infty}^{\infty}p(y_i|x_i,\sigma_{x_i},\sigma_{y_i},k,b)p(b|\mu,\sigma)\mathrm{d}b\\
&=\int_{-\infty}^{\infty}\frac{1}{\sqrt{2\pi(\sigma_{x_i}^2k^2+\sigma_{y_i}^2)}}\exp{\left(-\frac{(k x_i+b-y_i)^2}{2(\sigma_{x_i}^2k^2+\sigma_{y_i}^2)}\right)} \frac{1}{\sqrt{2\pi}\sigma}\exp{\left(-\frac{(b-\mu)^2}{2\sigma^2}\right)}\mathrm{d}b \\
&=\frac{1}{\sqrt{2\pi(\sigma_{x_i}^2k^2+\sigma_{y_i}^2+\sigma^2)}}\exp{\left(-\frac{(k x_i+\mu-y_i)^2}{2(\sigma_{x_i}^2k^2+\sigma_{y_i}^2+\sigma^2)}\right)},
\end{split}
\end{equation}
and we can write the data likelihood as
\begin{equation}\label{eq:likelihood}
p(\{y_i\}|\{x_i,\sigma_{x_i},\sigma_{y_i}\},k,\mu,\sigma) =\prod_{i=1}^{n}\frac{1}{\sqrt{2\pi(\sigma_{x_i}^2k^2+\sigma_{y_i}^2+\sigma^2)}}\exp{\left (-\frac{(k x_i+\mu-y_i)^2}{2(\sigma_{x_i}^2k^2+\sigma_{y_i}^2+\sigma^2)}\right)}.
\end{equation}

\begin{figure}
  \centering
  \includegraphics[width=\textwidth]{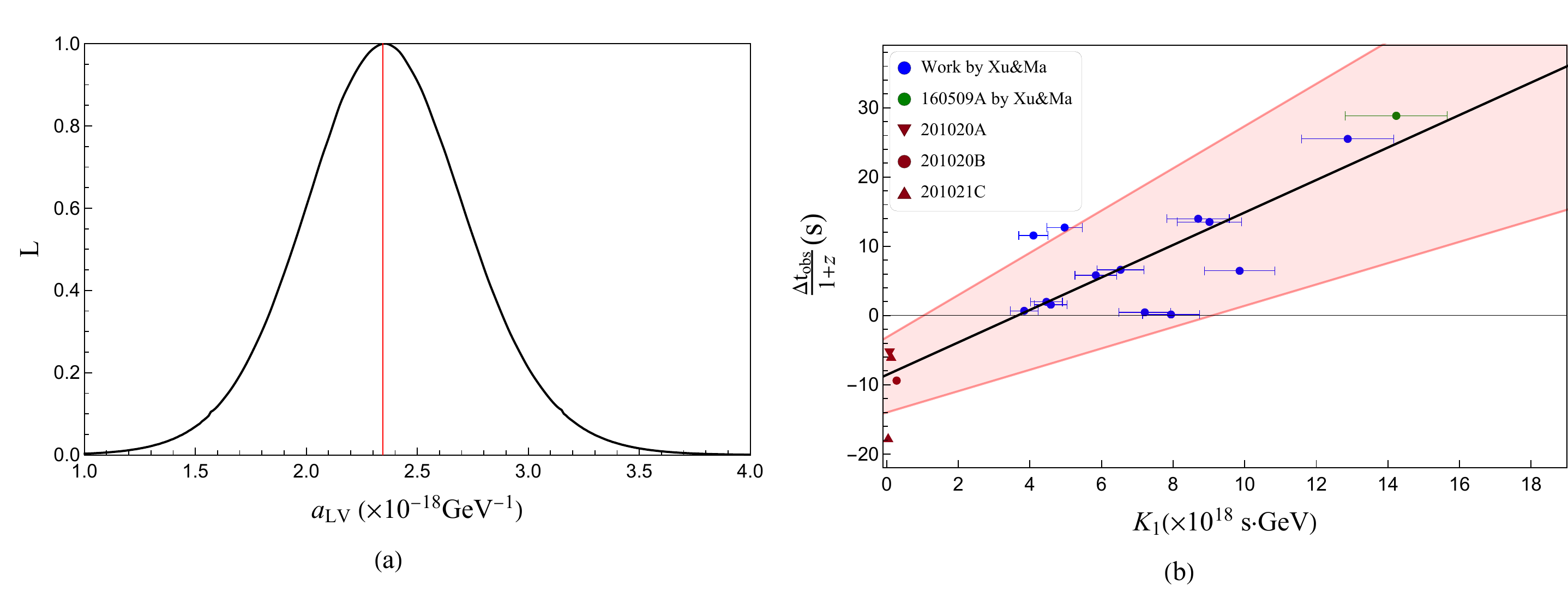}\\
  \caption{The likelihood function $L$ for the slope parameter~(left panel) and the plot for the fit of the data from Refs.~\cite{xu1,xu2}~(Xu\&Ma) and Table~\ref{tab:events}~(right panel). In the left panel~(a), the red vertical line represents the maximum likelihood estimation~(MLE) result for $a_{\rm LV}.$
  In the right panel (b), the black~(thick) straight line represents the fit result, and for the red~(thin) lines of the edge of the red~(shaded) area, the slopes are the edge of 95\% CL range for $a_{\rm LV}$
  and the intercepts are $\mu+\sigma$ and $\mu-\sigma$ while $\mu$ and $\sigma$ are the best fit parameters.\label{fig:all}}
\end{figure}

\begin{figure}
  \centering
  \includegraphics[width=\textwidth]{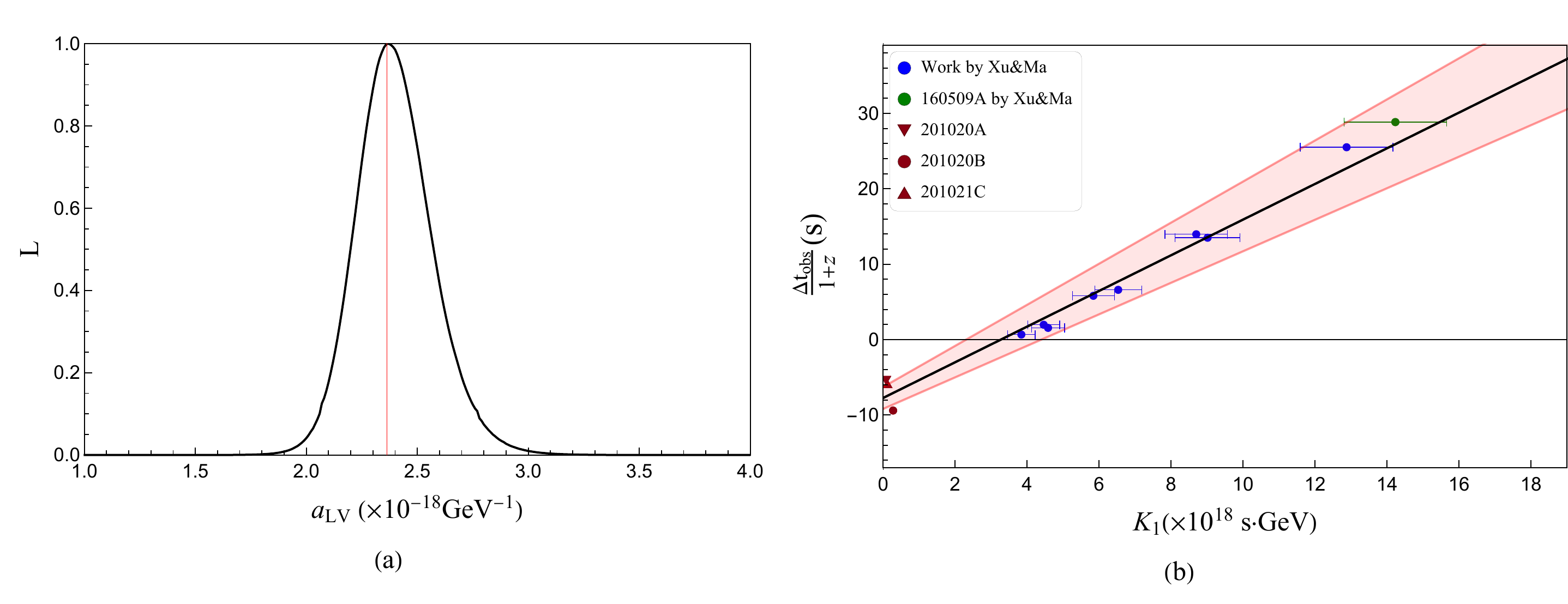}\\
  \caption{The likelihood function $L$ for the slope parameter~(left panel) and the plot for the fit of the data near the mainline from Refs.~\cite{xu1,xu2}~(Xu\&Ma) and Table~\ref{tab:events}~(right panel). In the left panel~(a), the red vertical line represents the maximum likelihood estimation~(MLE) result for $a_{\rm LV}.$
  In the right panel~(b), the black~(thick) straight line represents the fit result, and for the red~(thin) lines of the edge of the red~(shaded) area, the slopes are the edge of 95\% CL range for $a_{\rm LV}$
  and the intercepts are $\mu+\sigma$ and $\mu-\sigma$ while $\mu$ and $\sigma$ are the best fit parameters.\label{fig:main}}
\end{figure}

For the high-energy events with positive arrival time, we use the data of Refs.~\cite{xu1,xu2}. Then we add
the high-energy events with negative arrival time listed in Table~\ref{tab:events}. Considering the energy
resolution of LAT~\cite{LAT} (within 10\% uncertainty) and the uncertainties of the cosmological parameters, $\Delta t_{\rm{obs}}/(1+z)$ and $K_1$
can be calculated as shown in Table~\ref{tab:calc}. Since the time resolution for GBM and LAT is smaller than 10~$\mu s$~\cite{GBM_time,LAT_time} and it is much more smaller than the time scale in our data,
we assume that there is no error in $\Delta t_{\rm{obs}}$. Thus we
can write the likelihood function as
\begin{equation}\label{eq:likelihooddata}
L=C\exp{\left [-\frac{1}{2}\sum_{i=1}^{n} \left(\frac{\left(\frac{\Delta t_{{\rm obs}_i}}{1+z_i}-a_{\rm LV}K_i-\mu\right)^2}{\sigma^2+a_{\rm LV}^2\sigma_{k_i}^2}+\ln(\sigma^2+a_{\rm LV}^2\sigma_{k_i}^2)\right)\right ]},
\end{equation}
where $C$ is a constant, $a_{\rm LV}=1/E_{\rm LV,1}$, $K_i$ is $K_1$ of the $i$th data, $\mu$ and $\sigma$ are the mean and the standard deviation of $\Delta t_{\rm{in}}$.
The MLE result of the likelihood function for the data from Table~\ref{tab:calc} and Refs.~\cite{xu1,xu2} is
$a_{\rm LV}=2.35\times10^{-18}~{\rm GeV}^{-1}$, $\mu=-8.57~\rm s$ and $\sigma=5.43~\rm s$,
and the 95\% CL range for the slope is
$a_{\rm LV}=2.35_{-0.73}^{+0.83}\times10^{-18}~{\rm GeV}^{-1}$.
The likelihood function for the slope parameter and the plot for the fit are shown in Fig.~\ref{fig:all}.
The main contribution to uncertainties of $K_1$ is the energy resolution of LAT. The energies of the observed pre-burst events are too small compared
with those with positive arrival time, so we can hardly see the uncertainties in this figure. This result gives a limit on $E_{\rm LV,1}$ that $E_{\rm LV,1}=4.2_{-1.1}^{+1.9}\times10^{17}~{\rm GeV}$.
From Fig.~\ref{fig:all} we can see clearly that the three of the points in this work are near the main line of Refs.~\cite{xu1,xu2}, and we also choose the data near the mainline and fit them again.
The MLE result of the likelihood function for the data near the mainline is
$a_{\rm LV}=2.36\times10^{-18}~{\rm GeV}^{-1}$, $\mu=-7.73~\rm s$ and $\sigma=1.45~\rm s$,
and the 95\% CL range for the slope is
$a_{\rm LV}=2.36_{-0.27}^{+0.36}\times10^{-18}~{\rm GeV}^{-1}$. The likelihood function for the slope parameter and the plot for the fit are shown in Fig.~\ref{fig:main}.
Fig.~\ref{fig:main} suggests strong correlation between the data in Table~\ref{tab:calc} and Refs.~\cite{xu1,xu2}, and the result suggests that $E_{\rm LV,1}=4.23_{-0.56}^{+0.56}\times10^{17}~{\rm GeV}$.

\begin{threeparttable}[t]
  \begin{centering}
    \caption{Values of $\Delta t_{\rm{obs}}/(1+z)$ and $K_1$ of observed pre-burst events}
    \begin{tabular}{cp{50mm}<{\centering}cp{50mm}<{\centering}}
      \hline
      \hline
      GRB        & $\Delta t_{\rm{obs}}$(s) & $\Delta t_{\rm{obs}}/(1+z)$(s) & $K_1$($\times 10^{17} \rm{s}\cdot\rm{GeV}$) \\
      \hline
      201020A    & -21.08                   & -5.40                          &$0.81\pm0.08$  \\
      \hline
      201020B    & -16.95                   & -9.40                          & $2.77\pm0.28$                               \\
      \hline
      201021C(1) & -12.21                   & -5.90                          & $1.23\pm0.13$                               \\
      201021C(2) & -36.55                   & -17.66                         & $0.41\pm0.04$                                \\
      \hline
      \hline
    \end{tabular}%
    \label{tab:calc}%
  \end{centering}
\end{threeparttable}%
\\

%\begin{figure}[!h]
%  \centering
%  \subfloat[]{
%    \label{fig:1a}
%    \begin{minipage}{0.48\textwidth}
%      \centering
%      \includegraphics[width=\textwidth]{lhall.pdf}
%    \end{minipage}
%  }
%  \subfloat[]{
%    \label{fig:1b}
%    \begin{minipage}{0.52\textwidth}
%      \centering
%      \includegraphics[width=\textwidth]{fitall.pdf}
%    \end{minipage}
%  }

%\begin{figure}[!h]
%  \centering
%  \subfloat[]{
%    \label{fig:1a}
%    \begin{minipage}{0.48\textwidth}
%      \centering
%      \includegraphics[width=\textwidth]{lhmain.pdf}
%    \end{minipage}
%  }
%  \subfloat[]{
%    \label{fig:1b}
%    \begin{minipage}{0.52\textwidth}
%      \centering
%      \includegraphics[width=\textwidth]{fitmain.pdf}
%    \end{minipage}
%  }

We need to consider the probability of a large uncertainty around 5 seconds in $\Delta t_{\rm obs}$ due to the determination of the first
significant peak $t_{\rm low}$ of low energy photons. With an err bar of 5-seconds introduced for each $\Delta t_{\rm obs}$ in the analysis,
the MLE result corresponding to Fig.~\ref{fig:main} suggests that
$a_{\rm LV}=2.30\times10^{-18}~{\rm GeV}^{-1}$, $\mu=-7.05~\rm s$ and $\sigma=0.00~\rm s$.
We can understand the result of
$\sigma=0.00~\rm s$
in this way: the main difference between
whether we introduce an error $\sigma_{y_i}$ for every point or not is equivalent to adopt an effective $\tilde{\sigma}=\sqrt{\sigma^2+\sigma_{y_i}^2}$ or just $\sigma$ in Eq.~(\ref{eq:likelihooddata}). If the best fit for $\sigma$ of the data without $\sigma_{y_i}$ is bigger than $\sigma_{y_i}$, we can absorb the error of data into the parameter $\sigma$ of the normal distribution and get the result $\sigma=\sqrt{\tilde{\sigma}^2-\sigma_{y_i}^2}$. However in our previous fitting the result gives us $\sigma=1.45~\rm s$ and almost every error for the data $\sigma_{y_i}=5/(1+z_i)$ seconds is bigger than it, so it is reasonable to get the best estimation of $\sigma=0.00~\rm s$. The 95\% CL range for the parameters are
$a_{\rm LV}=2.30_{-0.30}^{+0.49}\times10^{-18}~{\rm GeV}^{-1}$ and
$\mu=-7.05_{-2.87}^{+2.01}~\rm s$,
and thus we get
$E_{\rm LV,1}= 4.34_{-0.76}^{+0.65}\times 10^{17} ~\rm GeV$.

\section{Discussion and Conclusion}

The result shown in Fig.\ref{fig:main} provides novel signals with significance. First, it supports the conclusion
of light speed variation first suggested in Ref.~\cite{xu1} and soon supported by a remarkable high energy event of GRB160509A in Ref.~\cite{xu2}. Second, it supports the conclusion of a
pre-burst stage of GRBs, and as suggested, at about 10 seconds before a gamma-ray burst at the source, there is a pre-burst stage of
high-energy photons with energy of multi-GeV scale. Third, although the condition on the observed pre-burst events is too
strict, we still find these events and make analysis on them and the result supports the previous conclusion. These events
might be regarded as background noises if one lacks of a pre-burst scenario, but analysis on them suggests that they are signals from the pre-burst stage of GRBs. So
we should treat observed pre-burst events with important significance as normal high energy photon events.

From another point of view, our analysis of multi-GeV photon events belongs to the catalog of looking for sharp peak structure behind the data. These multi-GeV photons from GRBs
are within a very large duration of one hundred seconds with different arrival times, but our analysis indicates that they come from the sources with a same intrinsic time. This implies that these multi-photon events on/near the mainline, if drawing an emission curve with time in the GRB source frame, correspond to a very sharp peak at -7.73~second with a very narrow width of only 2-3 seconds. So our result actually represents the finding of a significant sharp structure of multi-GeV photons from the Fermi GRB data.

In conclusion, we searched for high-energy photon events with earlier arrival time from GRBs with redshift recorded from FGST,
and found four pre-burst events of high energy photons from GRBs GRB~201020A, GRB~201020B and GRB~201021C.
Analysis on the observed pre-bursts events reveal that three of the observed pre-burst events from these GRBs
fall near the mainline that indicates a regularity of high energy photons. The result suggests
$E_{\rm LV,1}=4.23_{-0.56}^{+0.56}\times10^{17}~{\rm GeV}$ and the mean value of
$\Delta t_{\rm in}$ is $-7.73~\rm s$ with a width 2$\sigma=2.90$~s, which supports the earlier suggestion in Refs.\cite{xu1,xu2} for a light speed variation
$v(E)=c(1-E/E_{\mathrm{LV}})$ with $E_{\mathrm{LV}}=3.6\times10^{17}~\mathrm{ GeV}$ and a 10-second earlier
pre-burst stage of high energy photons of GRBs.
\\

%\section
\noindent
{\bf{Acknowledgements:}}
We thank the anonymous reviewer for the enlightening suggestions that have helped us to improve the quality of the analysis.  This work is supported by National Natural Science Foundation of China (Grant No.~12075003).




\vspace{1cm}

\begin{thebibliography}{99}
  \bibitem{method1}
 % G.~Amelino-Camelia {et al.},
 % Distance measurement and wave dispersion in a Liouville-string approach to quantum gravity.
  %  {Int.\ J.\ Mod.\ Phys.\ A} { 12} (1997) 607.
    G.~Amelino-Camelia, J.~R.~Ellis, N.~E.~Mavromatos, D.~V.~Nanopoulos,
Distance measurement and wave dispersion in a Liouville-string approach to quantum gravity.
{Int. J. Mod. Phys. A} {12} (1997) 607.
%{\it Int. J. Mod. Phys. A} {\bf 12} 607-624 (1997),

  \bibitem{method2}
  %G.~Amelino-Camelia {et al.},
 % Tests of quantum gravity from observations of gamma-ray bursts.
  %  { Nature} { 393} (1998) 763.
    G.~Amelino-Camelia, J.~R.~Ellis, N.~E.~Mavromatos, D.~V.~Nanopoulos, S.~Sarkar,
Tests of quantum gravity from observations of $\gamma$-ray bursts.
{Nature} {393} (1998) 763.

  \bibitem{formula}
  U.~Jacob, T.~Piran,
  Lorentz-violation-induced arrival delays of cosmological particles.
    { JCAP}  {0801} (2008) 031.


  \bibitem{pgb}
  K.~A.~Olive {et al.}, (Particle Data Group)
%  {Chinese Physics C }  {38} 090001 (2014).
  {Chinese Physics C }  {38} (2014) 090001.

\bibitem{Ellis}
%Ellis,~J.~R., Mavromatos,~N.~E., Nanopoulos,~D., Sakharov,~A.~S., \& Sarkisyan,~E.~K.~G.\
J.~R.~Ellis, N.~E.~Mavromatos, D.~Nanopoulos, A.~S.~Sakharov, E.~K.~G.~Sarkisyan,
%Robust limits on Lorentz violation from gamma-ray bursts.
Robust limits on Lorentz violation from gamma-ray bursts.
%{\it Astropart.\ Phys.\ } {\bf 25}, 402 (2006) [Corrigendum {29}, 158 (2008)].
{Astropart.\ Phys.\ } {25} (2006) 402-411. [Corrigendum {29} (2008) 158-159].

\bibitem{shaolijing}
%Shao,~L., Xiao,~Z. \& Ma,~B.-Q.\
L.~Shao, Z.~Xiao, B.-Q.~Ma,
%Lorentz violation from cosmological objects with very high energy photon emissions.
Lorentz violation from cosmological objects with very high energy photon emissions.
%{\it Astropart.\ Phys.\ } {\bf 33}, 312 (2010).
{Astropart.\ Phys.\ } {33} (2010) 312-315.

\bibitem{zhangshu}
%Zhang,~S., Ma,~B.-Q.\
S.~Zhang, B.-Q.~Ma,
%Lorentz violation from gamma-ray bursts.
Lorentz violation from gamma-ray bursts.
%{\it Astropart.\ Phys.\ } {\bf 61}, 108 (2015).
{Astropart.\ Phys.\ } {61} (2015) 108-112.


  \bibitem{xu1}
  H.~Xu,B.-Q.~Ma,
  Light speed variation from gamma-ray bursts.
    {Astropart. Phys.}  {82} (2016) 72.

  \bibitem{xu2}
  H.~Xu,B.-Q.~Ma,
  Light speed variation from gamma ray burst GRB 160509A.
    {Phys. Lett. B } {760} (2016) 602.
%    {Physics Letters B } {760} (2016) 602.

  \bibitem{LAT}
  W.~B.~Atwood {et al.},
  The Large Area Telescope on the Fermi Gamma-ray Space Telescope mission.
    { Astrophys.\ J.\ } {697} (2009) 1071.

  \bibitem{GBM}
  C.~Meegan, G.~Lichti, P.~N.~Bhat {et al.},
  The Fermi Gamma-Ray Burst Monitor.
    {Astrophys.\ J.\ } {702} (2009) 791.

  \bibitem{gbm_data}
  https://heasarc.gsfc.nasa.gov/FTP/fermi/data/gbm/triggers/

  \bibitem{lat_data}
  https://fermi.gsfc.nasa.gov/cgi-bin/ssc/LAT/LATDataQuery.cgi

  \bibitem{Liu}
  Y.~Liu, B.-Q.~ Ma,
  Light speed variation from gamma ray bursts: criteria for low energy photons.
    {Euro. Phys. J. C} {78} (10) 825.
  %  {The European Physical Journal C} {78} (10) 1.
  \bibitem{ScienceTool}
  https://fermi.gsfc.nasa.gov/ssc/data/analysis/scitools/overview.html

  \bibitem{201020Az}
  D.~A.~Kann, A.~de Ugarte Postigo, M.~Blazek, {et al.},
  GRB~201020A: Redshift from GTC/OSIRIS.
  {GCN Circ.\ } {28717} (2020).

  \bibitem{201020Bz}
  D.~A.~Kann, A.~de Ugarte Postigo, M.~Blazek, {et al.},
  GRB~201020B: Redshift from GTC/OSIRIS.
  {GCN Circ.\ } {28765} (2020).

  \bibitem{201021Cz}
  J.-B. Vielfaure, D. Xu, J. Palmerio, {et al.},
  GRB~201021C: VLT/X-shooter redshift.
  {GCN Circ.\ } {28739} (2020).



\bibitem{GBM_time}
https://fermi.gsfc.nasa.gov/ssc/data/analysis/documentation/Cicerone/Cicerone\_Introduction/GBM\_overview.html

\bibitem{LAT_time}
{https://fermi.gsfc.nasa.gov/ssc/data/analysis/documentation/Cicerone/Cicerone\_Introduction/LAT\_overview.html}

\bibitem{HESS}
A. Abramowski, et al. (H.E.S.S.),
Search for Lorentz Invariance breaking with a likelihood fit of the PKS 2155-304 flare data taken on MJD 53944.
Astropart. Phys. 34 (2011) 738.

\end{thebibliography}


\end{document}